\documentclass[useAMS,usenatbib]{mn2e}

\usepackage{multirow}
\usepackage{graphicx}
\usepackage{dcolumn}
\usepackage{bm}
\usepackage{url}

\title[Neptune Trojans distributions affected by planetary migration]{The effect of orbital damping during planet migration on the Inclination and Eccentricity Distributions of Neptune Trojans}
\author[Chen Yuan-Yuan, Ma Yuehua, Zheng Jiaqing]{Chen Yuan-Yuan$^{1}$\thanks{E-mail: chenyy@pmo.ac.cn}, Ma Yuehua$^{1}$\thanks{E-mail: yhma@pmo.ac.cn}, Zheng Jiaqing$^{2}$\\
$^{1}$Key Laboratory of Planetary Sciences, Purple Mountain Observatory, Chinese Academy of Sciences, Nanjing 210008, China\\
$^{2}$Tuorla Observatory, 215000 Piikki\"{o}, Finland}
\begin{document}

\date{Accepted 2015 ** **. Received 2015 ** **; in original form 2015 ** **}

\pagerange{\pageref{firstpage}--\pageref{lastpage}} \pubyear{2015}

\maketitle

\label{firstpage}

\begin{abstract}
{We explore planetary migration scenarios for formation of high inclination Neptune Trojans (NTs) and how they are affected by the planetary migration of Neptune and Uranus. If Neptune and Uranus's eccentricity and inclination were damped during planetary migration, then their eccentricities and inclinations were higher prior and during migration than their current values. Using test particle integrations we study the stability of primordial NTs, objects that were initially Trojans with Neptune prior to migration. We also study Trans-Neptunian objects captured into resonance with Neptune and becoming NTs during planet migration. We find that most primordial NTs were unstable and lost if eccentricity and inclination damping took place during planetary migration. With damping, secular resonances with Neptune can increase a low eccentricity and inclination population of Trans-Neptunian objects increasing the probability that they are captured into 1:1 resonance with Neptune, becoming high inclination NTs. We suggest that the resonant trapping scenario is a promising and more effective mechanism explaining the origin of NTs that is particularly effective if Uranus and Neptune experienced eccentricity and inclination damping during planetary migration.
}
\end{abstract}


\begin{keywords}
methods: numerical - celestial mechanics - Kuiper belt: general - minor planets, asteroids: general - planets and satellites: dynamical evolution and stability - planets and satellites: general.
\end{keywords}

\section{Introduction}

Trojans are planetesimals orbiting a central massive body near the stable Lagrange equilibrium points (L4 \& L5) of another massive body, which may be a planet if the central object is a star, or a moon if the central object is a planet. Here we focus on Trojans with a planet in orbit around a star. Trojans have orbital period similar to that of their host planet and so are said to be corotating with the planet or in 1:1 resonance with the planet. Trojan asteroids have been discovered corotating with Jupiter \citep{fernandez2003, grav2011}, Earth \citep{connors2011}, Mars \citep{fuente2013, scholl2005}, Uranus \citep{alexan2013} and Neptune \citep{sheppard2010, marzari2003}. As of Dec. 2015, 6,261 Jupiter Trojans have been identified \footnote{This number was provided by \url{http://www.minorplanetcenter.net/}}, and this is the largest group of Trojans in the solar system. There are also several Trojan moons corotating with Saturn.

12 Neptune Trojans (NTs) have been found in all, including 9 moving around the L4 Lagrange point and 3 around L5. Their orbital parameters are listed in Table \ref{data}. The orbital inclinations of 5 NTs are higher than $20^\circ$. \citet{sheppard2006} corrected for the observatory regional bias for the three NTs that they discovered to estimate the ratio of high- to low-inclination orbits of the L4 Trojans, finding a ratio of nearly 4:1. Recently, \cite{parker2015} improved upon correcting for the detection bias as a function of inclination, eccentricity, and libration amplitude distributions and estimated a standard deviation for the NT inclinations of $\sigma_i>11^\circ$ at greater than $95\%$ confidence level.

\begin{table*}
\label{data}
 \centering
 \begin{minipage}{140mm}
  \caption{Orbits of 12 observed Neptune Trojans at epoch JD = 2457000.5, except $2014 ~QP441$ at epoch JD=2456960.5,
  from {http://www.minorplanetcenter.net/.}}
  \begin{tabular}{@{}llrrrrrr@{}}
  \hline
Designation & $L_N$ & a(au) & e & i(deg) & $\Omega$(deg) & $\omega$(deg) & M(deg) \\
 \hline
2001 QR322  & $L_4$ & 30.2076 & 0.026603 & 1.323 & 151.665 & 158.144 & 73.899 \\
2004 KV18   & $L_5$ & 30.2020 & 0.187031 & 13.590 & 235.598 & 294.871 & 65.133 \\
2004 UP10   & $L_4$ & 30.1298 & 0.026943 & 1.435 & 34.759 & 7.255 & 346.073 \\
2005 TN53   & $L_4$ & 30.1268 & 0.067553 & 25.033 & 9.319 & 88.263 & 297.982 \\
2005 TO74   & $L_4$ & 30.1208 & 0.054846 & 5.258 & 169.448 & 305.309 & 278.746 \\
2006 RJ103  & $L_4$ & 30.0454 & 0.032237 & 8.161 & 120.956 & 24.915 & 254.381 \\
2007 VL305  & $L_4$ & 30.0791 & 0.064786 & 28.141 & 188.665 & 218.551 & 3.083 \\
2008 LC18   & $L_5$ & 29.9220 & 0.085442 & 27.575 & 88.518 & 9.366 & 176.119 \\
2011 HM102  & $L_5$ & 30.0587 & 0.079730 & 29.418 & 100.988 & 150.635 & 28.434 \\
2012 UV177  & $L_4$ & 30.0324 & 0.074148 & 20.819 & 265.726 & 204.568 & 292.934 \\
2014 QO441  & $L_4$ & 30.0893 & 0.104754 & 18.825 & 107.100 & 112.460 & 168.790 \\
2014 QP441  & $L_4$ & 30.0765 & 0.068423 & 19.395 & 96.618 & 3.194 & 297.887 \\
\hline
\end{tabular}
\end{minipage}
\end{table*}

The occurrence and high proportion of NTs with high inclinations should constrain dynamical formation mechanisms for NTs and so the orbital evolution of the giant planets. Nearly all researchers attribute the origin of high-inclination orbits of the NTs to a resonant capture process where planetesimals in a trans-Neptunian disk are captured into 1:1 resonance during an early epoch of planetary orbit instability or migration, similar to mechanisms for the capture of Jovian Trojans into 1:1 resonance \citep{morbidelli2005}. Using several long timescale numerical simulations of the current giant planet configuration, \citet{guan2012b} showed that low inclination NTs would be excited to high inclination in a very low efficiency. \citet{michtchenko2004} found that NTs were more dynamically stable when Neptune migrated at a slower rate (migration timescale $\tau = 10^6~yr$). They also found that NTs could escape resonance when secondary mean motion resonances between Uranus and Neptune perturbed Neptune. \citet{lithwick2005} discussed three origins of large NTs (radius $\sim100~km$) --pull-down capture, direct collision, and \textit{in situ} accretion-- and predicted that large NTs are likely to be located in a vertically thin disk. Using a series of simulations, \citet{lykawka2009, lykawka2011} studied both the transport of primordial NTs and the capture of new NTs from a debris disk. They found that objects captured into resonance during migration could account for the orbital characteristics of the present-day NTs. Orbits of individual NTs have also been studied to place constraints on their stability and dynamical evolution \citep{guan2012, horner2012a, horner2012b}.


\citet{nesvorny2009} investigated the capture process of the NTs within the exact context of the Nice model. In their simulations, gravitational interactions between planetesimals and the planets induce the planetary migration. 27,028 equal-mass planetesimals are placed into the primordial Trans-neptunian disk, at semi-major axes from 21AU to 35AU and before Jupiter and Saturn enter the 1:2 mean motion resonance. After migration, the simulations exhibited a dynamically exited planetesimal disk consistent with the $\sim4:1$ ratio of high- to low-inclination population of NTs. Their model can more realistically simulate the evolution history to some extent. While the 1:2 MMR of Jupiter and Saturn is a chaotic phase, most evolution results may be so different from the current configuration that only a few cases are effective.
%
Instead of computing planetesimal and planet gravitational interactions, \citet{parker2015} integrated the orbits of planetesimals while adding artificial forces to the planets causing them to migrate and damp in eccentricity. They set the characteristic timescale for eccentricity damping ($3\times10^5 ~yr$) to be smaller than the migration timescale ($10^6\sim10^7~yr$). \citet{parker2015} found that high inclinations of NTs arenot excited during the process of planetesimals captured but due to original high inclinations in trans-Neptunian disk. They also found that Planetesimals originally in high-inclination orbits are more easily captured into 1:1 resonance than planetesimals in low-inclination orbits \citep{nesvorny2009}.

In this paper, we study the role of eccentricity and inclination damping of the ice giants (i.e. Uranus and Neptune) during planetary migration. We investigate the effects of the orbital damping on two populations of NTs: 1) Objects transported with Neptune that were primordial NTs, captured into 1:1 resonance prior to migration; 2) Objects originating in a trans-Neptunian disk that are trapped into 1:1 resonance during planetary migration. A detail of Nice model is that the giant planets went through a set of resonant scattering events that excited their eccentricities and inclinations \citep{nesvorny2009}. Subsequently dynamical friction and scattering of the remaining planetesimals slowly reduced or damped the eccentricities and inclinations of planets. Here we study the effect of eccentricity and inclination damping during migration of Uranus and Neptune, as they are the primary perturbers of the NTs. We study how this damping affects the capture and evolution of NTs.

There are four sections in this manuscript. In Section 2, we introduce the numerical model. Section 3 provides the main results, which include the primordial NTs that are transferred by the migrating Neptune and the NTs that are trapped from the trans-Neptunian disk. The last part is the conclusions and discussions.

\section[]{Modeling planetary migration and orbital dissipation}

Each of our simulations includes the four giant planets of the solar system and one massless test particles (the planetesimal). All planets migrate radially with migration rate set with force per unit mass \citep{zhou2002, li2006}
\begin{equation}\label{drdt}
\Delta\ddot{\textbf{r}}=\frac{\hat{\textbf{v}}}{\tau}\bigg\{\sqrt{\frac{GM_\odot}{a_f}}
-\sqrt{\frac{GM_\odot}{a_i}}\bigg\}\exp\bigg(-\frac{t}{\tau}\bigg),
\end{equation}
where $a_f$ is the current semi-major axis of the planet and $a_i=a_f-\Delta a$ is the semi-major axis at the beginning of the simulation. We set the timescale $\tau=10^6$~yr and $\Delta a=-0.2,0.8,3.0,$ and $7.0$~AU for Jupiter, Saturn, Uranus and Neptune, respectively, following \citet{michtchenko2004}.

We mimic the dissipation induced by gravitational interactions with planetesimals by applying smooth eccentricity-damping and inclination-damping forces to Neptune and Uranus. Our applied force per unit mass
$\textbf{a}_e$ and $\textbf{a}_i$ are those used by \citet{cresswell2008},
\begin{equation}\label{ae}
\textbf{a}_e=-2\frac{(\textbf{v}\cdot\textbf{r})\textbf{r}}{r^2 t_e}
\end{equation}
and
\begin{equation}\label{ai}
\textbf{a}_i=-\frac{v_z}{t_i}\textbf{k},
\end{equation}
where $\textbf{k}$ is the unit vector in the \textit{z}-direction; $\textbf{r}$ and $\textbf{v}$ are the position and velocity vectors of the planets, respectively. Here $t_e$ and $t_i$ are the eccentricity and inclination damping timescales. We set $t_e=t_i=\tau=10^6$~yr in our simulations, based on the Nice model measurements by \citet{levison2008} and \citet{tsiganis2005}.

By substituting Equations (\ref{ae}) and (\ref{ai}) into Lagrange's equations, we find a rate of change of semi-major axis (migration rate)
\begin{equation}
\frac{da}{dt}=\frac{4a}{t_e}(\sqrt{1-e^2}-1),
\end{equation}
rate of change of eccentricity
\begin{equation}
\frac{de}{dt}=\frac{2(1-e^2)}{et_e}(\sqrt{1-e^2}-1),
\end{equation}
and inclination
\begin{equation}
\frac{di}{dt}=-\frac{\sin{i}}{2t_i}.
\end{equation}
The other orbital elements are not affected by these dissipative forces. As the migration rate ($\dot a$) depends on eccentricity, eccentricity damping affects the migration rate. With higher eccentricity $\dot a$ is lower. So that our simulations with and without eccentricity and inclination damping have the same migration rate at the beginning of the simulation, we have increased the final semi-major axis, $a_f$, of Uranus and Neptune for the simulations with damping, see Table \ref{initial}. Initial semi-major axes, set by the conditions at the end of the Nice model, are the same for both sets of simulations.


The initial eccentricities and inclinations of the four giant planets are set at representative values of their current mean values, except for the $e_0$ and $i_0$ values of Uranus and Neptune for the simulations with eccentricity and inclination damping. We used moderate initial values of $e_{N,0}=0.1$ and $i_{N,0}=10^\circ$, that were also used by \citet{wolff2012} and \citet{parker2015}. An even higher value of $i_{N,0}=20^o$ has been used by studies on Uranus's obliquity \citep{boue2010} and the irregular satellites of Uranus \citep{parisi2008}.

Particle and planet orbits are numerically integrated using the MERCURY code \citep{chambers1999}. Dissipation forces (Equations (\ref{drdt})-(\ref{ai})) are added as  non-gravitational forces to the planets. The hybrid symplectic integration algorithm is chosen to achieve moderate accuracy and increase the speed of calculation. For all simulations we used a time step of 200 days, which is approximately 1/20 of the orbital period of Jupiter. We simulate 1,000 small bodies in each individual integration run since self-gravity among planetesimals is neglected. 300-400 CPU-hours is required per 1,000 particle run, with the exact time depending on the different numbers of small bodies that are scattered out of the system during the integration.


\begin{table}
\label{initial}
\caption{Initial orbital elements of the four giant planets for our numerical integrations. Simulations are divided into two sets, those without eccentricity and inclination damping and those with eccentricity and inclination damping. The initial eccentricities and inclinations in the no-damping cases are those by \citet{murray1999}. The final semi-major axes, $a_f$, of Uranus and Neptune in the with-damping case have been adjusted so that the migration rates are similar at the beginning of the simulations. The other initial orbital elements (mean motion, argument of perihelion and longitude of the ascending node) are chosen arbitrarily.}
\begin{center} \footnotesize
\begin{tabular}{cccccc}
\hline\hline
Models & & Jupiter & Saturn & Uranus & Neptune \\
\hline
    & $a_i$ & 5.4 & 8.74 & 16.19 & 23.07 \\
\hline
no-damping & $a_f$ & 5.2 & 9.54 & 19.19 & 30.07 \\
    & $e_0$ & 0.048 & 0.054 & 0.047 & 0.009 \\
    & $i_0(deg)$ & 1.3 & 2.5 & 0.8 & 1.8 \\
\hline
with-damping & $a_f$ & 5.2 & 9.54 & 22.19 & 32.07 \\
    & $e_0$ & 0.048 & 0.054 & 0.1 & 0.1 \\
    & $i_0(deg)$ & 1.3 & 2.5 & 20 & 10 \\
\hline\hline
\end{tabular}
\end{center}
\end{table}
\vspace{0.0mm} {\footnotesize}\vspace{0.0mm}

\begin{figure}
\includegraphics[width=80mm,angle=0]{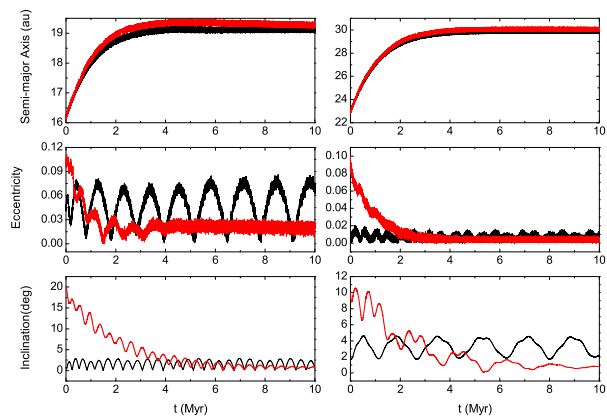}
  \caption{Comparison of the evolution of orbital elements of Uranus (left) and Neptune (right) in the no-damping and with-damping cases. The black lines represent the no-damping case, and the red lines the with-damping case. From top to bottom, the panels are semi-major axis, eccentricity and inclination versus time.}
  \label{planet}
\end{figure}

In Figure \ref{planet} we show the evolution of semi-major axis, eccentricity and inclination of Uranus and Neptune from two of our integrations. The semi-major axis evolution for integrations with and without eccentricity and inclination damping are similar and the final planet eccentricities and inclinations approximate their current values.

\section{Numerical Simulations and Results}

Test particles are initially placed near the stable L4 and L5 Lagrange points of Neptune. These particles begin the simulation corotating with Neptune and are used to study stability of NTs during migration and the inclination evolution of bodies that were captured into resonance prior to migration. We also run simulations with test particle initially placed outside of Neptune's orbit. These are used to study how particles are captured into the 1:1 resonance from the primordial Trans-Neptunian belt.

\subsection{Examples of orbital evolution of particles initially corotating with Neptune}


\begin{figure}
\includegraphics[width=80mm,angle=0]{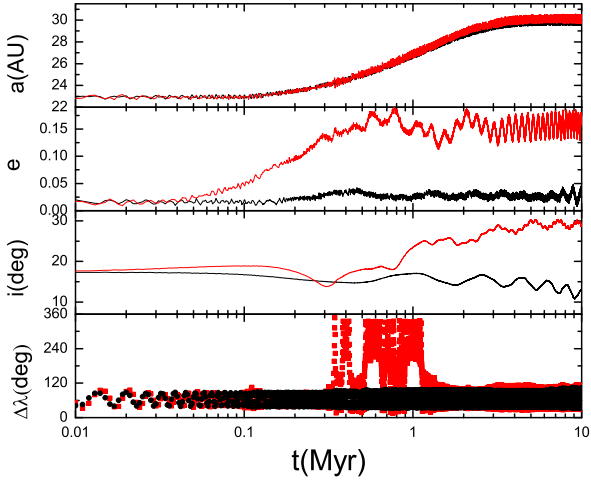}
  \caption{Evolution of the orbital elements of a test particle initially corotating with Neptune and comparing evolution with and without eccentricity and inclination damping of Neptune and Uranus. From top to bottom we plot semi-major axis, eccentricity, inclination and the Neptune-particle 1:1 resonant angle $\Delta\lambda=\lambda_p-\lambda_N$ versus time. The black lines in these panels shows the no-damping case, and the red lines the with-damping case.}
  \label{example1}
\end{figure}

\begin{figure}
\includegraphics[width=80mm,angle=0]{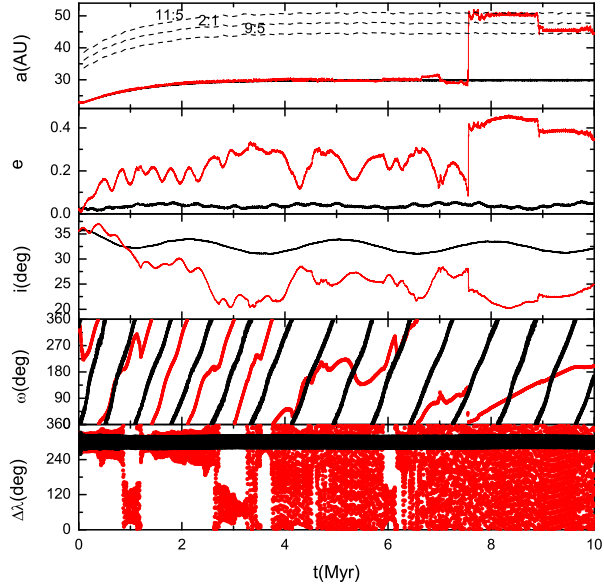}
  \caption{The same as Fig. \ref{example1}, except for another test particle. The evolutions of the argument of pericentre $\omega$ is also shown so that it is possible to see Kozai oscillation.}
  \label{example2}
\end{figure}

We select a corotating test particle initially near the L4 Lagrange point, and contrast the orbital element evolution with and without eccentricity and inclination damping of Neptune and Uranus. In Fig. \ref{example1} we plot semi-major axis, eccentricity, inclination and Neptune-particle 1:1 resonant angle $\Delta\lambda=\lambda_p-\lambda_N$ versus time.
Here $\lambda$ and $\lambda_p$ are the mean longitudes of the particle and Neptune. With damping, both the eccentricity and inclination of the test particle increase.
The increase in particle inclination and eccentricity occurs within 1~Myr when Neptune has a higher inclination and eccentricity in the simulation with damping than in the simulation without damping (see Figure \ref{planet}). We attribute these increases to secular resonances with Neptune that are stronger when Neptune's inclination and eccentricity
are higher. The excitation in eccentricity and inclination affect the particle's stability. The orbit converts from a tadpole orbit to a horseshoe orbit between $0.3$~Myr and $1.2$~Myr as seen by excursions of $\Delta \lambda$ during this interval. In contrast, the particle remains near the L4 point (with $\Delta \lambda$ librating about $60^\circ$) throughout the 10~Myr evolution in the no-damping case. Without eccentricity and inclination damping, the particle's eccentricity and inclination remain near their initial values. This example illustrates a dramatic difference in the orbital element evolution of the test particle that is due to eccentricity and inclination damping.

In Fig. \ref{example2} we illustrate the evolution of a different test particle placed initially near the L5 Lagrange point. With Neptune damping, the particle hops between the L4 and L5 points until it escapes from the 1:1 resonance with Neptune at approximately $4$ ~Myr. Then Kozai oscillation \citep{kozai1962, chen2013} begins, and the particle maintains the same semi-major axis with Neptune. At $\sim7.5$~Myr, the test particle is scattered into a more distant orbit. The jumps in semi-major axis and eccentricity imply that the scattering is caused by high-order mean-motion resonances with Neptune. The orbital variation arise from the commensurability between the Neptune-Trojan 1:1 mean-motion resonance and Uranus-Neptune mean-motion near-resonance \citep{michtchenko2004}, or secular resonances of giant planets \citep{dvorak2008, zhou2009}.

\subsection{The stability of primordial NTs}

We now study a distribution of primordial NTs, or test particles begun near L4 or L5 Lagrange points with Neptune. We consider how planetary migration could have affected the distribution of their orbital elements.
%
A distribution of test particles is created with the following properties
\begin{equation}
a=a_{N,0},
\end{equation}
so that the particle is corotating with Neptune. Particles are chosen with mean longitude in two groups
\begin{eqnarray}
L4: |\triangle\lambda-60^\circ |<20^\circ, \nonumber\\
L5: |\triangle\lambda-300^\circ |<20^\circ,
\end{eqnarray}
and a uniform distribution used to select initial values for $\lambda$. Particles eccentricities and inclinations are chosen
\begin{eqnarray}
e: (0, 0.4) ~ {\rm uniform ~distribution}, \nonumber\\
i: (0^\circ, 40^\circ) ~{\rm uniform ~distribution}.
\end{eqnarray}
We desire our initial particle distribution to be stable so we first run integrations including only gravitational forces with the four giant planets, neglecting migration and e/i-damping. After 100,000 years of integration we chose 10,000 stable particles from the L4 region and the same number from the L5 region.

\begin{figure}
\centering
\includegraphics[width=80mm,angle=0]{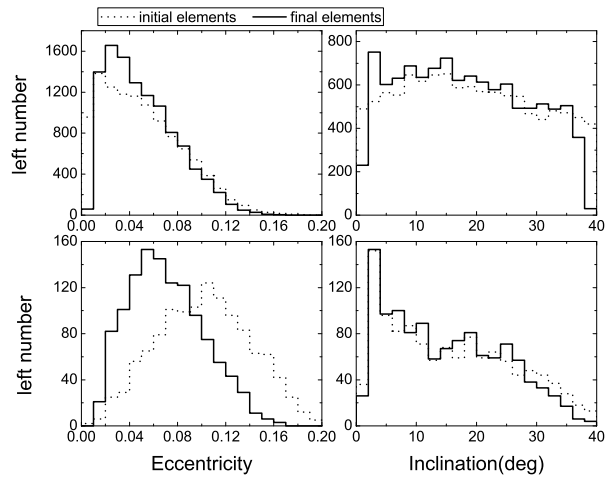}
\caption{Distribution of eccentricities and inclinations of the surviving primordial NTs before and after planet migration, in the no-damping case (top panels) and in the with-damping case (bottom panels). The dotted lines show the distributions of the remaining NTs before planets migrate but after 100,000 years of integration with the 4 giant planets using initial orbital elements listed in Table \ref{initial}. \label{distri11}}
\end{figure}

We then ran numerical integrations for $10^7$~yr now including orbital migration, with and without eccentricity and inclination damping. After these integrations we measure the number of remaining NTs (objects corating with Neptune) by counting the number of particles with  $|a-a_N|<2~AU$ and $\Delta\lambda\in(0^\circ, 120^\circ)\cup(240^\circ, 360^\circ)$.
The numbers of remaining NTs in the no-damping and with-damping cases are listed in Table 3, and their orbital element distributions are shown in Fig. \ref{distri11} as solid lines. For the case with e/i damping of Uranus and Neptune, the primordial NTs are more unstable, and the number of remaining NTs is approximately 1/8 of that in the no-damping case (1,198 compared to 10,876).
A comparison of the inclination distributions (right two panels of Fig. \ref{distri11}) shows that high inclination NTs are particularly unstable and likely to be lost. There is a shift of the distribution of eccentricities between the two cases (comparing the dotted line with the solid line in the left-side bottom panel), which we are not sure about the probable reason.

\begin{table}
\label{data1}
\caption{Numbers of remaining NTs at the end of planet migration in the no-damping and with-damping cases for 20,000 test particles, half of which are initially located near L4 and the other half of which are initially near L5. The ratio of the number in L4 to that in L5 is $f_{45}=N(L4)/N(L5)$. Objects are classified as stable using criteria based on the current planetary configuration of the solar system in \citet{zhou2011}.}
\begin{center} \footnotesize
\begin{tabular}{ccccc}
\hline\hline
  & The total & L4 & L5 & $f_{45}$ \\
\hline
  no-damping case & 10,876 & 5,486 & 5,390 & 1.02 \\
  stable    &  6923 & 3478 & 3445 & 1.01 \\
\hline
  with-damping case & 1,198 & 910 & 288 & 3.16 \\
  stable    &  357 &  167 & 190 & 0.88 \\
\hline\hline
\end{tabular}
\end{center}
\end{table}

\begin{figure}
\centering
\includegraphics[width=80mm,angle=0]{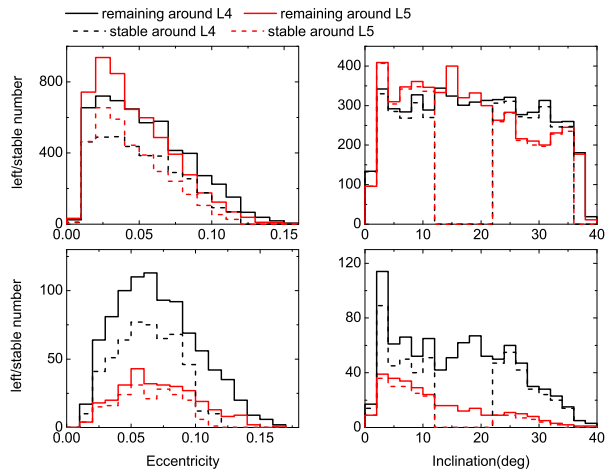}
\caption{Distribution of eccentricities and inclinations of the remaining primordial NTs after planet migration, in the no-damping case (top panels) and in the with-damping case (the bottom panels). The black lines denote the distributions of NTs around L4, and the red lines denote those around L5. The dashed lines bracket stable NTs using the criteria based on the current  configuration of the solar system by \citet{zhou2011}. \label{distri12}}
\end{figure}

We also compare the L4 eccentricity and inclination distributions to those at L5 in Fig. \ref{distri12}. Objects are classified as stable using the criterion by \citet{zhou2011}, who found that in the current solar system planetary configuration, there are three stable inclination regions ($(0^\circ, 12^\circ), (22^\circ, 36^\circ), (51^\circ, 59^\circ)$), with upper limits on eccentricity of $0.1, 0.12$ and $0.04$, respectively. Two of these regions are shown in Fig. \ref{ei} by gray areas. Of the objects remaining after orbital integration, the number of stable objects in L4 and L5 regions is also listed in Table ~3 and drawn in Fig. \ref{distri12} by dashed lines. The large value in Table 3 for the with-damping value of the ratio of objects (both stable and unstable) in L4 compared to L5, $f_{45}$, is transient as the excess objects are primarily unstable. The ratio of estimated stable objects in L4 compared to L5, $f_{45}$, is close to 1 with reasonable statistical fluctuations.

Fig. \ref{distri11} shows that the ratio of high- to low-inclination orbits is lower than the observed value of 4:1. We have modified the initial eccentricity and inclination distributions of the primordial NTs and find similar distributions, so the discrepancy in the ratio (between observed and shown in Fig. \ref{distri11}) is not resolved by beginning with a wider inclination distribution.

\subsection{NTs captured during orbital migration}

NTs could have originally been objects in the Trans-Neptunian belt that were captured into 1:1 resonance with Neptune during planetary migration. Although resonant capture is likely to be inefficient, this mechanism might be more efficient at high than low inclination \citep{lykawka2009}.

\begin{figure}
\centering
\includegraphics[width=80mm,angle=0]{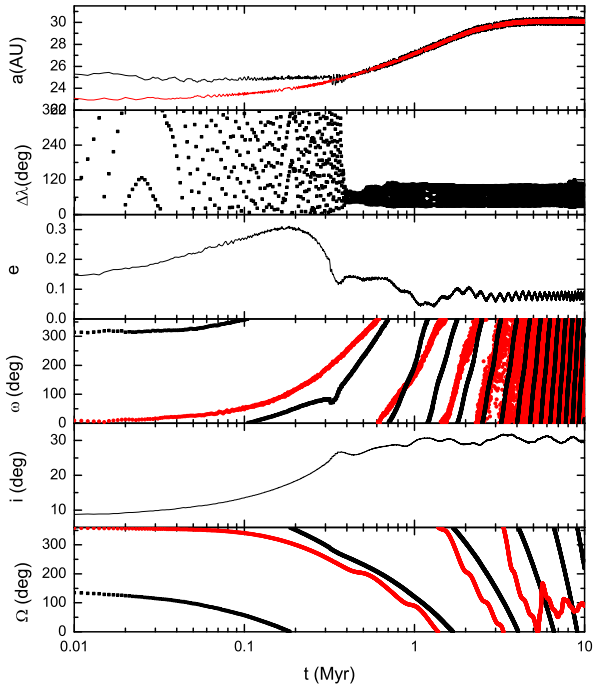}
\caption{Shown as black lines are the semi-major axis, 1:1 resonance angle with Neptune, eccentricity, argument of perihelion $\omega$, inclination and the longitude of ascending node, $\Omega$, of a test particle, as a function time, integrated  with eccentricity and inclination damping. The red lines show the orbital  elements of Neptune. The initial orbital elements of the particle are $a=25.078$~AU, $e=0.115, i=8.495^\circ, \omega=309^\circ, \Omega=146^\circ$, and mean anomaly $M=101^\circ$. The object is captured into the L4 area at about 0.35~Myr. Before resonance capture, its eccentricity and inclination are excited by  secular resonances with Neptune. \label{example3}}
\end{figure}

In Fig. \ref{example3} we illustrate the orbital evolution of an object that is captured into the L4 region. Both the eccentricity and inclination are excited by secular resonances with Neptune before the test particle is captured at approximately 35,000 yr. Between $0.2-0.25$~Myr, the inclination later increases to even higher values while eccentricity decreases. Because the argument of perihelion librates during this time we infer that the inclination increase is caused by Kozai resonance. According to the statistical studies by \citet{nesvorny2009} and \citet{parker2015}, higher eccentricity and inclination planetesimals are more easily trapped as NTs. So stirring-up or heating processes (secular resonances and the Kozai mechanisms) can indirectly increase the probability that objects are captured as NTs.

To estimate the likelihood that Trans-Neptunian objects are trapped as NTs we carried out a series of numerical integrations. Test particles are initially placed outside the orbit of primordial Neptune, from 23.7~AU ($=a_{N,0}+a_{N,Hill}$) to 32~AU, covering the range of semi-major axis covered by Neptune during its migration. Here $a_{N,Hill}$ is Neptune's Hill radius. Trapped Trojans are identified if they lie within $|a-a_N|<2~AU$ and resonant angle $\Delta\lambda\in(0^\circ, 120^\circ)\cup(240^\circ, 360^\circ)$ during the last 1~Myr of the integration. Integrations were run for $10^7$~yr.

\begin{figure}
\centering
\includegraphics[width=80mm,angle=0]{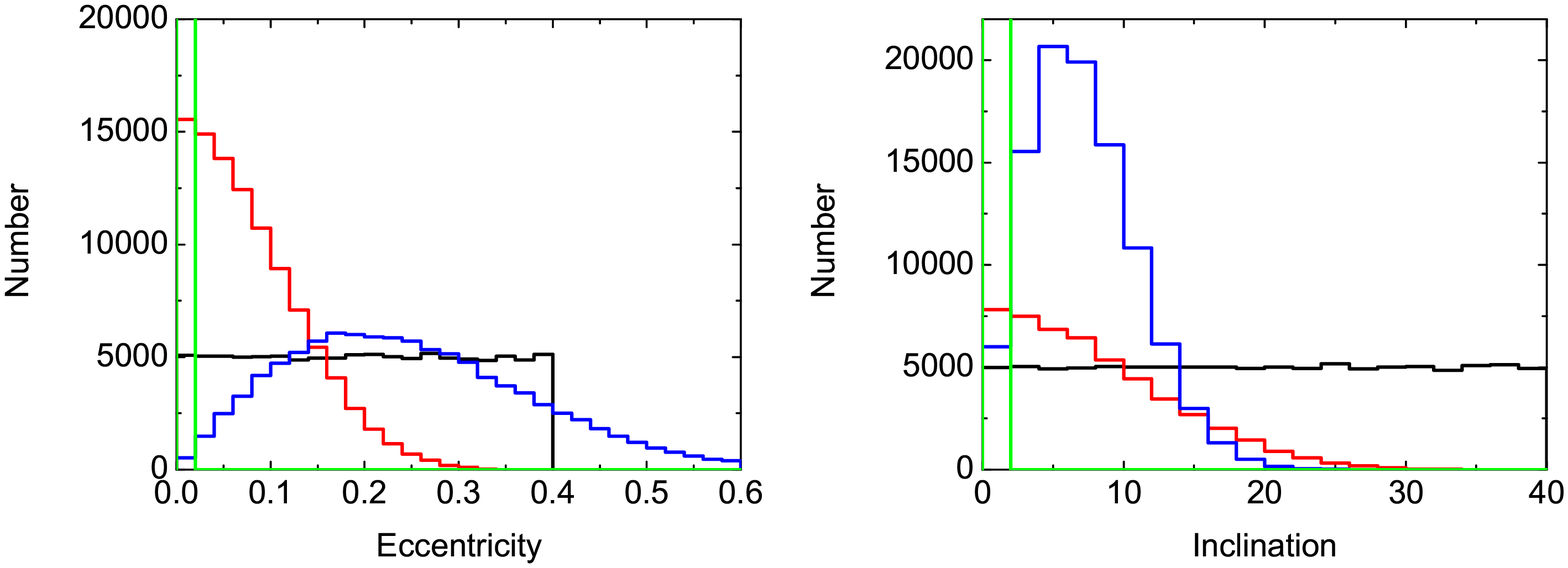}
\caption{Initial distributions of eccentricities and inclinations of the test particles in the trans-Neptunian disk. The black line shows the uniform distribution, the red line the normal distribution, and the blue and green lines show hot and cold $e,i$ Rayleigh distributions, respectively. \label{initialdis}}
\end{figure}

Initial eccentricities and inclinations were drawn from four different distributions:
\begin{enumerate}
\item A uniform distribution:  $e \in (0, 0.4), i \in (0^\circ, 40^\circ)$.
\item  Normal distributions with means and standard deviations for eccentricity and inclination
$\mu_e=0, \sigma_e=0.1, \mu_i=0, \sigma_i=10^\circ$.  When drawing from the normal distribution
we discarded negative values.
\item hot Rayleigh distributions with means
$ \mu_e=0.2, \mu_i=0.1$ radians $(\sim5.6^\circ)$.
\item  cold Rayleigh distributions with means
$ \mu_e=0.001, \mu_i=0.0005$ radians $(\sim0.03^\circ)$
\end{enumerate}
The last two distributions are consistent with those estimated for the hot and cold primordial Kuiper Belt by \citet{hahn2005}. The four initial distributions of eccentricities and inclinations are shown in Fig. \ref{initialdis}.

\begin{figure}
\centering
\includegraphics[width=80mm,angle=0]{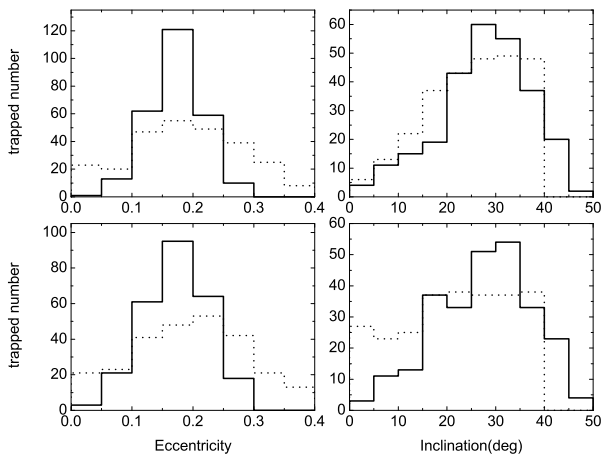}
\caption{Distribution of eccentricities and inclinations of captured NTs after planet migration, in the no-damping case (top panels) and in the with-damping case (bottom panels). The simulations shown here were done with an initial uniform eccentricity and inclination distribution. Neglecting objects that were not captured as NTs, the dotted lines denote the initial e/i distributions of only the objects identified as captured NTs. \label{distri21}
}
\end{figure}

For initial uniform particle distributions (in eccentricity and inclination), Fig. \ref{distri21} shows eccentricity and inclination distributions of the captured NTs after $10^7$~yr of evolution. The number of captured NTs is listed as the first line in Table~4. A comparison of Fig. \ref{distri21} with Fig. \ref{distri11} shows that for both no-damping and with-damping migration cases, the fraction of captured NTs with high inclinations ($20^\circ \sim 40^\circ$) compared to those with low inclinations ($i<20^\circ$) is larger than that of seen in our primordial NT simulations. We find that NTs with high inclinations are more likely to originate from the resonant capture mechanism than be primordial NTs, which confirms the previous study by \citet{lykawka2009}. The large high inclination fraction may be due to excitation by secular resonances with Neptune, the same process that causes instability in high inclination primordial NTs. As we drew from an initial uniform distribution, there are initially similar fractions of high- and low-inclination orbits, which causes that the secular resonances do not increase the fraction of high-inclination orbits, and so the numbers and final distributions in the no-damping and with-damping cases are similar. We also find that, the with-damping case gives more trapped Trojans from low inclination objects compared with the no-damping case (comparing the dot line in the right-side two panels in Fig. \ref{distri21}). This shows that more initially low-inclination planetesimals are excited to high-inclination orbits and captured into NTs in the with-damping case. 

\begin{figure}
\centering
\includegraphics[width=80mm,angle=0]{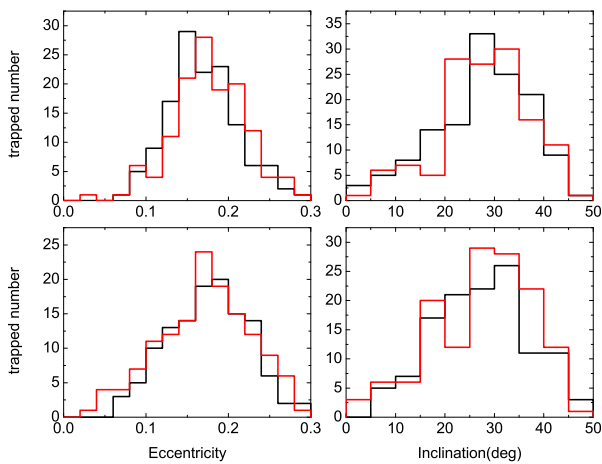}
\caption{
Distribution of eccentricities and inclinations of captured NTs after planet migration around L4 and L5, in the no-damping case (top panels) and in the with-damping case (the bottom panels). Here initial particle eccentricities and inclinations were drawn from a uniform distribution. The black lines denote the distributions of NTs around L4, and the red lines denote those around L5.
\label{distri22}}
\end{figure}

Fig. \ref{distri22} is a comparison of the number of trapped NTs in L4 and L5, with an initial uniform $e,i$ distributions. We find there are similar numbers of captured NTs at L4 and L5, and that the orbital element distributions are also similar to each other.

The numbers of trapped NTs for the different initial inclination and eccentricity distributions are listed in Table ~4. The capture efficiency is higher in with-damping cases than no-damping cases for all initial distributions except the uniform one. Furthermore, in with-damping case, the trapping efficiency is higher for the cold Rayleigh distribution than the hot Rayleigh distribution (263 vs 174), whereas in the no-damping cases, the situation is the opposite, i.e., the trapped NT number in the hot Rayleigh case is much higher in the with-damping case (109 vs 29). These phenomenons may also be attributed to secular resonance with Neptune before the test particles are captured into 1:1 resonance (Fig. \ref{example3}). The fraction of low-inclination objects is larger than the fraction of high-inclination objects in the initial three distributions, except for the uniform one (Fig. \ref{initialdis}). Thus, secular excitation in with-damping cases increases the proportion of high-inclination orbits by turning low-inclination into high-inclination ones, which can enhance the capture efficiency \citep{nesvorny2009,parker2015}. Specifically, secular excitation in with-damping cases boosts the capture planetesimals from the cold Rayleigh distribution condition by increasing the fraction of high-inclination orbits, meanwhile it suppresses the fraction captured from the hot Rayleigh distribution by causing instability in the high-inclination objects.

\begin{table}
\label{trapn}
\caption{Numbers of captured NTs at the end of planetary migration for different initial inclination and eccentricity distributions. The two numbers in the brackets represent the numbers of objects in L4 and L5 regions. Objects are classified as stable using the criterion by \citet{zhou2011} that is based on the current planetary configuration. }
\begin{center} \footnotesize
\begin{tabular}{ccc}
\hline\hline
Models & no-damping cases & with-damping cases \\
\hline
Uniform & 266(134,132) & 262(123,139) \\
stable & 12(5,7)    & 23(11,12)  \\
\hline
Normal & 124(61,63) & 278(123,155) \\
stable & 9(4,5)     & 18(7,11)   \\
\hline
hot Rayleigh & 109(59,50) & 174(82,92) \\
stable & 3(3,0)     &  9(6,3)   \\
\hline
cold Rayleigh & 29(11,18) & 263(134,129) \\
stable & 2(1,1)     &  14(9,5)  \\
\hline\hline
\end{tabular}
\end{center}
\end{table}

\begin{figure}
\centering
\includegraphics[width=80mm,angle=0]{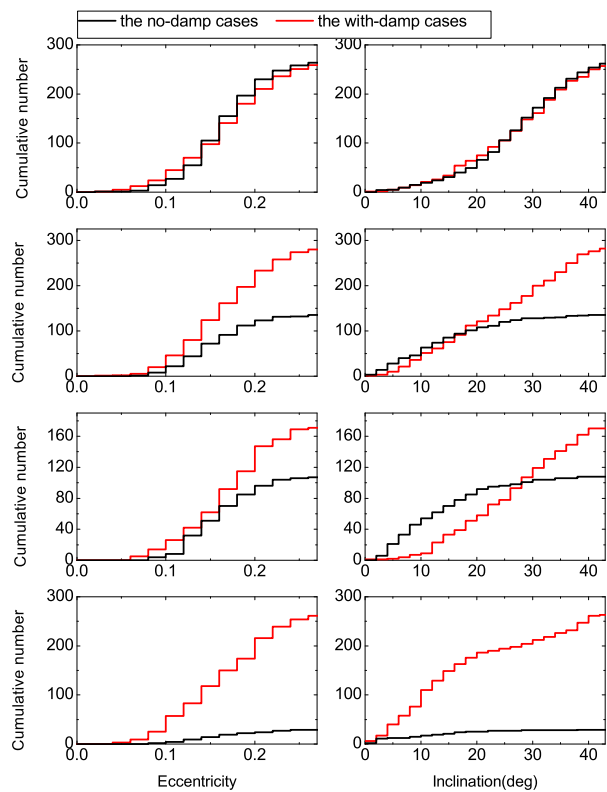}
\caption{Cumulative distributions of eccentricities and inclinations of captured NTs for the four different sets of initial inclination and eccentricity distributions. The black and red histograms show the no-damping cases and the with-damping cases, respectively. From top to bottom, the initial distributions are uniform, normal, hot Rayleigh or cold Raleigh as enumerated in section 3.3. The NTs are included in the histogram if they remain near a Lagrange point during the last $10^6$~yr of the $10^7$~yr integration.  \label{distri4}}
\end{figure}

We show in Fig. \ref{distri4} the cumulative distributions of final eccentricities and inclinations of trapped NTs for the different initial $e,i$ distributions.
Orbits all concentrate on the high eccentricity (inclination) part, as the situation shown in the uniform distribution (Fig. \ref{distri21}).

\subsection{Comparison with observed data in $e-i$ space}

\begin{figure}
\centering
\includegraphics[width=80mm,angle=0]{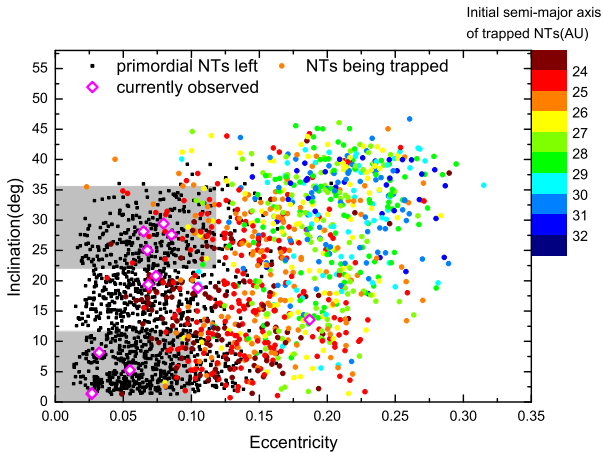}
\caption{Eccentricities versus inclinations of the primordial remaining and captured NTs at the end of all integrations. The primordial NTs are shown as black dots. Trapped NTs are shown as colored dots with color showing their semi-major axis when they were captured into resonance. The magenta diamonds denote the observed NTs. The wine and navy colors represent objects captured into resonance at a semi-major axis less than 24~AU and larger than 32~AU, respectively. Two grey areas show stable regions computed by \citet{zhou2011}. \label{ei}}
\end{figure}

NTs with high eccentricities and high inclinations are unstable dynamically under the current configuration of the solar system \citep{zhou2011}, consequently observed NTs with high inclinations have moderately low eccentricity ($e<0.1$ ~for ~$i>20^\circ$). To compare our integrated eccentricity and inclination distributions with the observed distribution, we plot eccentricity versus inclination for all remaining and captured NTs in the with-damping cases (Fig. \ref{ei}). On this plot, stable regions computed by \citet{zhou2011} are shown as rectangular gray areas. Though the number of captured high-inclination objects is larger than the number of captured low-inclination objects, most of the captured NTs are actually located in dynamically unstable regions, lying outside these gray areas.

The $e,i$ distribution shown in Fig. \ref{ei} also has indications of secular resonance with Neptune. The color bar on the righthand side of Fig. \ref{ei} shows the semi-major axis of planetesimals at the time when they were captured into the 1:1 resonance with Neptune. The blue end of the color bar shows planetesimals that were trapped from the outer part of the trans-Neptunian disk. These experienced a long period of time when they were excited by secular resonances with Neptune before they were captured. Thus, most of them have high eccentricities and inclinations and are located on the top and righthand region of Fig. \ref{ei}.

\section{Discussion and conclusion}\label{discuss}

We have investigated how the orbital element distributions of NTs are affected by planetary migration, specifically on how orbital dissipation of Uranus and Neptune while they migrated affects the distributions. We find that orbital dissipation or damping of eccentricity and inclination of Uranus and Neptune during migration plays a crucial role in affecting the  orbital distributions of NTs, for both primordial and captured NTs populations. For primordial NTs, the orbital dissipation causes instability, reducing the surviving fraction, from 0.54 to 0.06. The proportions of the remaining NTs with high inclinations ($20^\circ-40^\circ$) (compared to low inclinations) also decreases (Fig. \ref{distri11}). For trapped NTs, secular resonances with Neptune are important due to the initial higher eccentricity and inclination of Neptune during migration. These resonances excite inclinations and eccentricities in the trans-Neptune disk objects before they are captured into 1:1 resonance with Neptune. Firstly, in a dynamically cold initial trans-Neptune disks objects initially at low inclination are excited by secular resonances and then dominate
the population of high inclination captured NTs after migration. The trapping efficiency is enhanced by $224\%$ for an initial normal $e,i$ distributions, $160\%$ in the hot Rayleigh initial distribution ($\mu_e=0.2,\mu_i=0.1$~radian), and $907\%$ in the cold Rayleigh distribution condition ($\mu_e=0.001,\mu_i=0.0005$~radian), respectively where we are comparing simulations with-damping to those without. Secondly, the capture efficiency from the cold trans-Neptunian disk (cold Rayleigh distribution case, 263/100,000) exceeds that of the stirred-up disk (hot Rayleigh distribution case, 174/100,000) after considering the orbital dissipation of ice giants (i.e., with-damping cases). Thus, the orbital dissipation of Uranus and Neptune due to dynamical friction of planetesimal disks must be considered in future studies of the origins of the NT population.

The captured NTs population contains a larger number of objects at high eccentricity and high inclination than in our simulated primordial NT population, both in the with-damping case and in the no-damping case (Fig. \ref{distri21}). Though the capture efficiency for the NTs trapped from the primordial trans-Neptune disk is much lower than the proportion of surviving primordial NTs, it is still possible that the ratio of high-inclination to low-inclination orbit of NTs approaches the observed value 4:1, because the trans-Neptune disk originally had many more planetesimals than were originally present as part of a primordial NT population.

We have compared the numbers and orbital element distributions of NTs around the two Lagrange points, L4 and L5, with Neptune. The orbital distributions of the NTs around L4 and L5 are similar for both the primordial and trapped NT populations. The capture efficiencies are also similar. There is a transient asymmetry at the end of the planetary migration for the primoridal NTs in the with-damping case (Table ~3). Our estimated number of stable trapped NTs is too small to make meaningful statistical conclusions.

Since the final planetary architecture in our simulations is not entirely consistent with the current architecture of the solar system, we do not continue the calculations after $10^7$~yr. There are currently planetesimals passing through L4 and L5 regions with Neptune and collisions continue to change the number of trapped NTs. The number of NTs present at the end of our simulations at $10^7$ years should be different from present today. Furthermore, most of our simulated trapped NTs with high inclinations also have high eccentricities (see also \citet{lykawka2011}), and these are unstable if integrated to a longer time. How trapped high-eccentricities and high-inclinations NTs remain stable on long time scales remains a problem for future study.

\section*{Acknowledgments}

We would like to thank the referee for his (or her) helpful comments and suggestions, which have helped us to improve, clarify and complete the manuscript.  We thank Prof. Mauri Valtonen, Prof. Alice Quillen, Dr. SongHu Wang, and Dr. Lei Feng for their great help checking the paper and suggestions helping us complete the paper. This work is supported by the National Natural Science Foundation of China (Grant No. 11403107, 10933004, 11573075), the Natural Science Foundation of Jiangsu Province (Grant No. BK20141045), and the Minor Planet Foundation of the Purple Mountain Observatory.

\label{lastpage}


\begin{thebibliography}{99}
\bibitem[\protect\citeauthoryear{Alexandersen et al.}{2013}]{alexan2013}Alexandersen, M., Gladman, B., Greenstreet, S., et al., 2013, Science, 341, 994
\bibitem[\protect\citeauthoryear{Boue \& Laskar}{2010}]{boue2010}Boue, G., \& Laskar, J., 2010, ApJL, 712, L44
\bibitem[\protect\citeauthoryear{Chambers}{1999}]{chambers1999}Chambers, J.E., 1999, MNRAS, 304, 793
\bibitem[\protect\citeauthoryear{Chen et al.}{2013}]{chen2013}Chen, Y.-Y., Liu, H.-G., Zhao, G., \& Zhou, J.-L., 2013, ApJ, 769, 26
\bibitem[\protect\citeauthoryear{Chiang \& Lithwick}{2005}]{lithwick2005} Chiang, E.I., \& Lithwick, Y., 2005, ApJ, 628, 520
\bibitem[\protect\citeauthoryear{Connors et al.}{2011}]{connors2011}Connors, M., Wiegert, P., \& Christian, V., 2011, Nature, 475, 481
\bibitem[\protect\citeauthoryear{Cresswell \& Nelson}{2008}]{cresswell2008} Cresswell, P., \& Nelson, R.P., 2008, A\&A, 482, 677
\bibitem[\protect\citeauthoryear{de la Fuente Marcos \& de la Fuente Marcos}{2013}]{fuente2013}de la Fuente Marcos, C., \& de la Fuente Marcos, R., 2013, MNRASL, 432, 31
\bibitem[\protect\citeauthoryear{Dvorak et al.}{2008}]{dvorak2008} Dvorak, R., Lhotka, Ch., \& Schwarz, R., 2008, CMDA, 102, 97
\bibitem[\protect\citeauthoryear{Fernandez et al.}{2003}]{fernandez2003}Fernandez, Y.R., Sheppard, S.S., \& Jewitt, D.C., 2003, AJ, 126, 1563
\bibitem[\protect\citeauthoryear{Grav et al.}{2011}]{grav2011}Grav, T., Mainzer, A.K., Bauer, J., et al. 2011, ApJ, 742, 40
\bibitem[\protect\citeauthoryear{Guan et al.}{2012}]{guan2012}Guan, P., Zhou, L.-Y., \& Li, J., 2012, RAA, 12, 1549
\bibitem[\protect\citeauthoryear{Guan}{2012}]{guan2012b}Guan, P., 2012, The orbits of Neptune Trojans. Included by Nanjing University Library.
\bibitem[\protect\citeauthoryear{Hahn \& Malhotra}{2005}]{hahn2005} Hahn, J.M., \& Malhotra, R., 2005, AJ, 130, 2392
\bibitem[\protect\citeauthoryear{Horner et al.}{2012}]{horner2012a} Horner, J., Lykawka, P.S., Bannister, M.T., \& Francis, P., 2012, MNRAS, 422, 2145
\bibitem[\protect\citeauthoryear{Horner \& Lykawka}{2012}]{horner2012b} Horner, J., \& Lykawka, P.S., 2012, MNRAS, 426, 159
\bibitem[\protect\citeauthoryear{Kortenkamp et al.}{2004}]{michtchenko2004} Kortenkamp, S.J., Malhotra, R., \& Michtchenko, T., 2004, Icarus, 167, 347
\bibitem[\protect\citeauthoryear{Kozai}{1962}]{kozai1962}Kozai, Y., 1962, AJ, 67, 591
\bibitem[\protect\citeauthoryear{Levison et al.}{2008}]{levison2008}Levison, H.F., et al., 2008, Icarus, 196, 258
\bibitem[\protect\citeauthoryear{Li et al.}{2006}]{li2006} Li, J., Zhou, L.-Y., \& Sun, Y.-S., 2006, Chin. J. Astron. Astrophys. 6, 588
\bibitem[\protect\citeauthoryear{Lykawka et al.}{2009}]{lykawka2009} Lykawka, P.S., Horner, J., Jones, B.W., \& Mukai, T., 2009, MNRAS, 398, 1715
\bibitem[\protect\citeauthoryear{Lykawka et al.}{2011}]{lykawka2011} Lykawka, P.S., Horner, J., Jones, B.W., \& Mukai, T., 2011, MNRAS, 412, 537
\bibitem[\protect\citeauthoryear{Marzari et al.}{2003}]{marzari2003}Marzari, F., Tricarico, P., \& Scholl, H., 2003, A\&A, 410, 725
\bibitem[\protect\citeauthoryear{Morbidelli et al.}{2005}]{morbidelli2005}Morbidelli, A., Levison, H.F., Tsiganis, K., \& Gomes, R., 2005, Nature, 435, 462
\bibitem[\protect\citeauthoryear{Murray \& Dermott}{1999}]{murray1999} Murray, C.D., \& Dermott, S.F. 1999, Solar system dynamics (Cambridge Univ. Press)
\bibitem[\protect\citeauthoryear{Nesvorny \& Vokrouhlicky}{2009}]{nesvorny2009} Nesvorny, D., \& Vokrouhlicky, D., 2009, AJ, 137, 5003
\bibitem[\protect\citeauthoryear{Parisi et al.}{2008}]{parisi2008}Parisi, M.G., Carraro, G., Maris, M., \& Brunini, A., 2008, A\&A, 482, 657
\bibitem[\protect\citeauthoryear{Parker}{2015}]{parker2015} Parker, A.H., 2015, Icarus, 247, 112
\bibitem[\protect\citeauthoryear{Scholl et al.}{2005}]{scholl2005}Scholl, H., Marzari, F., \& Tricarico, P. 2005, Icarus, 175, 397
\bibitem[\protect\citeauthoryear{Sheppard \& Trujillo}{2006}]{sheppard2006}Sheppard, S.S., \& Trujillo, C.A., 2006, Science, 313, 511
\bibitem[\protect\citeauthoryear{Sheppard \& Trujillo}{2010}]{sheppard2010}Sheppard, S.S., \& Trujillo, C.A., 2010, Science, 329, 1304
\bibitem[\protect\citeauthoryear{Tsiganis et al.}{2005}]{tsiganis2005}Tsiganis, K., Gomes, R., Morbidelli, A., \& Levison, H.F., 2005, Nature, 435, 459
\bibitem[\protect\citeauthoryear{Wolff et al.}{2012}]{wolff2012}Wolff, S., Dawson, R., \& Murray-Clay, R.A., 2012, ApJ, 746, 171
\bibitem[\protect\citeauthoryear{Zhou et al.}{2002}]{zhou2002} Zhou, L.-Y., Sun, Y.-S., Zhou, J.-L., Zheng, J.-Q., \& Valtonen, M., 2002, MNRAS, 336, 520
\bibitem[\protect\citeauthoryear{Zhou et al.}{2009}]{zhou2009} Zhou, L.-Y., Dvorak, R., \& Sun, Y.-S., 2009, MNRAS, 398, 1217
\bibitem[\protect\citeauthoryear{Zhou et al.}{2011}]{zhou2011}Zhou L.-Y., Dvorak, R., \& Sun, Y.-S. 2011, MNRAS, 410, 1849
\end{thebibliography}
\end{document}